# Beyond Nutrition Labels: How Analogical Reasoning Shapes Synthetic Media Disclosure Design

Claire R. Leibowicz

Oxford Internet Institute, University of Oxford, claire.leibowicz@oii.ox.ac.uk

As synthetic media proliferates, AI policymakers and practitioners have increasingly turned to disclosures—signals describing how media has been created or modified by AI—to help audiences evaluate media credibility. While there is a growing body of research on user interpretations, the upstream decision-making processes that affect users remain underexplored. This study therefore examines how AI policymakers and practitioners design synthetic media disclosures under complex sociotechnical constraints. Drawing on 23 expert interviews and 13 case studies from organizations participating in the Partnership on AI's Synthetic Media Framework, analysis identifies key disclosure goals, including process transparency and harm reduction, and two central tensions that emerge when pursuing those goals: normativity versus neutrality and proactivity versus precision. Findings highlight the role of analogical reasoning, from nutrition labels to Prop 65 warnings, in managing, but not resolving tensions. Ultimately, this study emphasizes the need for scholarship focused on AI transparency decision-makers and their use of analogical reasoning to support audiences encountering media in the AI age.

In 2024, Meta faced public backlash. Users had noticed the platform's "Made with AI" media transparency labels were appearing on images they believed had not, in fact, been made with AI (Mehta, 2024a; Mehta, 2024b). AI-assisted editing tools were to blame, as creators had used them for minor edits that nonetheless triggered AI disclosures. In response, Meta acknowledged that their labels "weren't always aligned with people's expectations, and didn't always provide enough context," before changing their labels and pledging to reevaluate their transparency policies (Bickert, 2024).

Meta is not alone in navigating the changing media landscape. AI policymakers and practitioners—a stakeholder cohort I describe broadly to include those working on policies, norms, standards, regulations, and technological interventions both within and beyond government—have focused for years on transparency as the solution to synthetic media's harms to the information ecosystem (Waddell, 2019; Gazis & Becket, 2019). From OpenAI to the California State Assembly, the BBC to the EU, synthetic media policies have converged on direct disclosures (OpenAI, 2022; Heikkilä, 2024; Gedye, 2025; BBC, 2025), signals for conveying whether a piece of media is AI-generated or AI-modified, like content labels, visible watermarks, and disclosure fields (e.g., disclaimers and warning statements) (Partnership on AI, 2023; Chandra, et al., 2024). In doing so, policymakers and practitioners may build "a more transparent ecosystem where users can tell whether digital media has been made by humans, AI, or some combination of both" (Riccardi, 2025). Yet, empirical studies paint a conflicting picture of how users interpret disclosures (Sanderson et al., 2025; Saltz et al., 2021a; Google, 2024), and varied stakeholders can hold different ideas about what counts as effective disclosure in the first place (Wittenberg et al, 2025; Saltz et al., 2021b).

While existing research has begun examining user interpretations—a crucial dimension of media literacy in the AI age—we know far less about how upstream stakeholders who implement these systems conceptualize disclosure's purpose and navigate design constraints. These stakeholders play a defining role in the media environments that users encounter: their judgments about what counts as "AI-generated," how that should be conveyed, and why it matters determine the very forms of transparency users experience. This paper therefore examines how policymakers and practitioners navigate these disclosure challenges—an essential yet understudied dimension of media transparency that serves audiences in the AI age.

## Literature Review

### Conceptualizing Disclosure Effectiveness

To situate this inquiry, it is helpful to examine how disclosure effectiveness has been conceptualized in prior research. Designing and evaluating disclosures is challenging, largely because researchers and platforms measure impact and define effectiveness differently. Wittenberg and colleagues (2024) help make sense of two common metrics, suggesting one involves whether disclosures "[communicate] to viewers the process by which a given piece of content was created or edited." This is often related to another goal: "diminishing the likelihood that content misleads or deceives its viewers." While the first goal may contribute to the second, it does not guarantee it; a disclosure might clarify how media was produced without preventing misinterpretation, or it might affect trust without conveying details about creation. This complicates both cross-study evaluation and choices facing policymakers and practitioners.

In this paper, I propose treating both goals as contributing to a broader aim: supporting audiences' ability to make effective media credibility judgments. While terms such as authenticity and accuracy and trust, despite their semantic nuances, are often used interchangeably, when studying AI policy, credibility serves as a high-level, multi-dimensional construct consisting of many of these other elements, a framing I borrow from Feng and colleagues (2023) and Hilligoss and Rieh (2008). This framing is particularly useful when studying transparency policy, as it's ultimately these credibility judgments that mitigate or exacerbate informational harms from synthetic media.

### Audience Research on the Impact of Synthetic Media Disclosure

Even when clear conceptual frameworks are present, empirical evidence reveals a stark disconnect between what stakeholders hope disclosures will accomplish and how audiences respond to them. This disconnect becomes particularly complicated when we consider that public opinion research that policymakers and practitioners rely on to make decisions often supports AI disclosures, suggesting that audiences desire greater transparency to "keep pace with rapid advances in generative AI technologies and usage" (Carpenter, 2024). For example, a Canadian survey of online harms from 2024 highlighted that 64 percent of respondents wanted online platforms to label synthetic media (Lockhart et al., 2024), and Meta's public opinion research from 13 countries also highlighted how 82 percent favored warning labels for AI-generated content featuring people saying things they did not say (Bickert, 2024). These findings position disclosure as a policy solution responsive to public sentiment, but research on the actual effects of labels reveals disclosures with a more complicated impact.

Beyond surveys, human-computer interaction experiments present modest evidence that providing audiences with signals about how content was edited or created supports media credibility assessments (Sanderson et al., 2025). For instance, Feng and colleagues (2023) found that visual disclosures linking to contextual provenance increased distrust in deceptive media; however, they also led respondents to doubt credible media. In contrast, Sanderson and colleagues (2025) found no such effect of provenance labels on credibility assessments. Other studies emphasize the limits of relying on neutral, technical descriptions of how a given piece of media was edited or created to obliquely gesture at media credibility (Bereskin, 2023). Collectively, these studies suggest that disclosures can shape audience judgments, but not in straightforward ways.

Broader digital media research reinforces these complexities (Bui et al., 2024; Saltz et al., 2021). The "implied truth effect" shows that warning labels on misleading information may lead to audiences assuming unlabeled media is credible (Pennycook et al., 2020). A similar phenomenon occurs with AI disclosures; an "implied authenticity effect" has been observed when audiences interpret unlabeled content as authentic simply because its origins are unknown (Sanderson et al., 2025; Google, 2024). Google (2025) observed this phenomenon, reporting that 35 percent of participants were highly trusting of unlabeled content after viewing labeled content, compared to 26 percent who saw unlabeled content in isolation.

Continued user research is essential to better understand the impacts of AI transparency and improve disclosure design. Yet, policymakers and practitioners are already deploying transparency systems that shape audience perception, even in the absence of conclusive user understanding. Investigating how these stakeholders make transparency decisions is therefore vital for designing disclosures that support media credibility in an age of synthetic media. Every disclosure interface—from a "Made with AI" label to elaborate provenance displays—represents a translation of policy intent into user experience, yet we know little about the institutional constraints and concepts guiding these translations.

**Analogical Reasoning in AI Transparency Design**

Given the limits of user research, policymakers and practitioners frequently invoke established transparency frameworks from other domains—analogical reasoning that shapes both effectiveness conceptualizations and design choices. Policy scholars have long recognized how concepts from other domains, like "generative metaphors," can support construction of policy problems and solutions (Schön, 1993; Schön & Rein, 1994). Consider how the nutrition label has become a recurrent conceptual tool in AI transparency overall and synthetic media disclosure specifically, appearing in projects from the Data Nutrition Project's dataset "ingredients" framework (Holland et al., 2018) to Content Credentials' "digital nutrition label for content" (Adobe, 2025). Decades of nutrition labeling research highlight key insights—like standardization, placement and prominence, and the risks of information overload—that can inform synthetic media disclosure design and evaluation. Beyond mere metaphors, AI stakeholders are selecting analogies that map functional relationships from one domain (e.g., food labeling) onto another (e.g., synthetic media disclosure).

In synthetic media disclosure, analogies act as frames, coordinating understanding and action across the policy-to-public pipeline. While framing theory encompasses multiple traditions (Dewulf et al., 2009), Entman's (1993) approach proves especially valuable for disclosure communication because it illuminates how stakeholders "select some aspects of a perceived reality and make them more salient...to promote a particular problem definition, causal interpretation, moral evaluation, and/or treatment recommendation" and since it originates in the media context (p. 52). Applied to synthetic media, analogical frames operate at two crucial levels: they organize how stakeholders develop AI transparency requirements, and they embed in disclosure designs that shape audience assessments.

This duality makes analogical reasoning particularly important for communication scholars studying media credibility in the AI age. When stakeholders adopt the nutrition frame, for instance, they implicitly define synthetic media as a consumer information problem (users need to know "ingredients" to make healthy choices), diagnose harm as resulting from absent disclosure (like hidden sugars causing health problems), assert moral claims about transparency and empowerment (right to know), and advocate specific remedies centered on standardized labeling (mandatory disclosure requirements). Consumers then encounter the results of such framing when navigating the media environment. Thus, such analogies provide both rhetorical power and empirical precedents for disclosure development amidst uncertain user research and complex sociotechnical forces.

## Motivation

Despite the centrality of practitioner and policymaker choices to AI disclosure outcomes, research has not examined how decision-makers navigate translation from policy goals to disclosure design. Doing so is particularly consequential for communication scholarship, since AI disclosures represent a site where technical affordances, institutional incentives, visual and media literacy, and audience interpretation converge. The design elements practitioners choose—whether to label content as "AI-generated" versus "Manipulated Media" or where to position labels—encode assumptions about visual communication, information processing, and trust that may fundamentally shape media credibility in the AI age. This paper therefore studies the policymakers and practitioners enacting synthetic media disclosure and grappling with what information to provide, how to present it, and how to balance competing goals in service of media credibility. Specifically, this paper seeks to answer:

- **RQ1:** How do stakeholders designing synthetic media disclosures conceptualize “effective” disclosure?
- **RQ2:** What analogical reasoning patterns do stakeholders employ when translating disclosure goals into communication design decisions, and how do these analogies enable and constrain implementation?

## Methods

Qualitative methods supported in-depth assessment of how those developing synthetic media disclosure practices understand and design them. Three main research phases took place: 1) expert interviews with 13 **AI Practitioners**— media, industry, and civil society stakeholders who have worked on synthetic media disclosure 2) expert interviews with 10 **AI Policymakers**— government policymakers who have contributed to synthetic media disclosure policies, and 3) archival analysis of 13 real-world case studies on synthetic media disclosure. These research phases were iterative rather than strictly sequential. AI practitioners came from many of the same organizations that wrote the real-world implementation cases, and insights from initial interviews guided closer case examination. This interaction illuminated both individual stakeholder reasoning and institutional implementation practices, thereby strengthening triangulation, contextualization, and insight into the constraints and trade-offs of synthetic media disclosure.

### Positioning Partnership on AI as Research Foundation

13 AI practitioners and 13[1] real-world case studies were drawn from organizations that participated in Partnership on AI's (PAI) Responsible Practices for Synthetic Media, a notable example of cross-sector synthetic media policy. Founded in 2016, PAI is one of the most mature and institutionally diverse responsible AI organizations and has worked on synthetic media governance since 2018 (Fjeld et al., 2020), making it a useful basis for studying disclosure design. Founded by Amazon, Facebook, Google, DeepMind, Microsoft, IBM, and later Apple, PAI now has over 140 global partner organizations from across sectors, including the ACLU and Creative Commons (Markoff, 2016).

In 2023, the organization released the Synthetic Media Framework, a voluntary technology policy focused on responsible synthetic media creation and distribution (Leibowicz & Cardona, 2024a). Disclosure is emphasized, suggesting stakeholders "disclose when the media [they] have created or introduced includes synthetic elements, especially when failure to know about synthesis changes the way the content is perceived." It provides high-level guidance for effectiveness metrics, suggesting that organizations "aim to disclose in a manner that mitigates speculation about content, strives toward resilience to manipulation or forgery, is accurately applied, and also, when necessary, communicates uncertainty without furthering speculation" (Partnership on AI, 2023).

The Framework has formal support from 18 institutions: Adobe, BBC, Bumble, CBC, Code for Africa, D-iD, Google, Stanford HAI, Meedan, Meta, Microsoft, OpenAI, Respeecher, Synthesia, Thorn, TikTok, Truepic, and WITNESS, representing established and start-up technology players, civil society, and media organizations. Notably, from 2024-2025, each organization submitted a transparency case study to "remedy the typical lack of accountability" for voluntary frameworks and ensure institutions "offer transparency" that could "deepen adherence to [PAI's] practices and principles" (Leibowicz & Cardona, 2024). Thus, PAI serves as a rich venue for studying real-world decision making on synthetic media disclosure.

### Interview Participants and Sampling

23 individual experts were interviewed in 60-minute, in-depth, semi-structured interviews (**Table 1).** Expert interviews enabled, as Meuser and Nagel (1991) have written, "privileged access to information" from individuals responsible for the "planning and provision of problem solutions," which is specifically relevant to the goal of understanding disclosure as a remedy for synthetic media's risks (Pfadenhauer, 2009). Importantly, the interviewer was also a known practitioner in synthetic media, ensuring that experts perceive them as a competent conversation partner, a prerequisite to empirically useful expert interviews (Honer, 1993 as cited in Pradenhauer, 2009).

Table 1: Distribution of Interviewees and Sectors

| **Sector** | **Respondents** |
|---|---|
| Civil Society | PAI1, PAI3, PAI10 |
| Industry | PAI2, PAI4, PAI6, PAI8, PAI11, PAI12, PAI13 |
| Media | PAI5, PAI7, PAI 9 |
| Government Policy | POL1 - POL10 |

#### *AI Practitioner Participants*

Purposive sampling identified 13 AI practitioners drawn from civil society, industry, and media participants involved in PAI's synthetic media disclosure work. Several stakeholders declined to participate due to employer constraints. The final cohort included three civil society, seven industry, and three media participants who contributed to disclosure designs while working across product, policy, research, news, and advocacy functions and a range of technical and social impact expertise.

[1] That both the practitioner interview and case study sample were 13 was coincidental, not intentional.

### ***AI Policymaker Participants***

Ten policymakers were recruited through snowball sampling, resulting in six from North America and four from Europe. Like AI practitioner participants, they came from a variety of disciplinary backgrounds and worked on voluntary guidance, legislation, and technical standards. Supplementing PAI interviewees ensured inclusion of a key stakeholder group thinking about disclosure notably absent from PAI work. While policymakers could be drawn from a global sample, I focused on those in Europe and North America; this was partially due to snowball sampling and expert access realities, but these two jurisdictions are also regulatory trendsetters in AI policymaking and map on to PAI stakeholder demographics (Engler, 2023), allowing a more coherent analysis alongside the PAI data.

## **Interview Protocol**

Interview questions were developed based on themes that emerged during autoethnographic reflection on the Synthetic Media Framework process. Both practitioners and policymakers answered questions exploring how they understand the effectiveness of synthetic media disclosure, how AI-generated media should be labeled for audiences, and what barriers must be overcome to support audiences. These questions complemented those in the case studies and established codes that could also be applied to case analysis.

One interview section focused on a disclosure elicitation task to probe the role of analogical reasoning in synthetic media disclosure design and implementation. Stakeholders were asked to think aloud about disclosure examples from other fields that they view as positive or negative models for synthetic media transparency design. This exercise surfaced 36 disclosure analogs informing how practitioners and policymakers frame choices about disclosure design.

## **Case Study Selection**

From the PAI case study collection, 13 cases were selected for archival analysis based on their focus on direct disclosure. PAI provided two case templates: a broad template (Partnership on AI, 2024a) and one specifically designed for "an underexplored area of synthetic media governance: direct disclosure" (Leibowicz & Cardona, 2024b). Code for Africa, Google, Meedan, Meta, Microsoft, Stanford HAI, Thorn, and Truepic used the disclosure-specific template, and OpenAI, TikTok, WITNESS, and the BBC wrote disclosure-focused cases from the broad template, making them equally relevant to analyze (**Table 2**) (Partnership on AI, 2024b). The disclosure template contains 31 questions exploring key themes: how organizations define direct disclosure, real-world implementation examples, perspectives on broader industry approaches, and the role of media literacy education (Partnership on AI, 2024b). The template specifically examines organizations' disclosure goals, including their relationship to accuracy and harm reduction, how media literacy complements artifact-level disclosures, and how research shapes disclosure practices.

Table 2: Case Study Organizations and Titles

| Organization | Title |
|---|---|
| Code for Africa | How an AI-manipulated video caused harm during South African elections — An analysis by digital democracy nonprofit Code for Africa |
| Google | How Google's research informed its approach to direct disclosure |
| Meedan | How an investigation in South Asia uncovered harmful synthetic media — An analysis by technology nonprofit Meedan |
| Meta | Meta updated its approach to direct disclosure based on user feedback |
| Microsoft | Microsoft and LinkedIn gave users detailed context about media on the professional networking platform |
| Stanford HAI | Direct disclosure has limited impact on AI-generated Child Sexual Abuse Material |
| Thorn | Mitigating the risk of generative AI models creating Child Sexual Abuse Materials – an analysis by child safety nonprofit Thorn |
| Truepic | Truepic used disclosures to help authenticate cultural heritage imagery in conflict zones |
| Adobe | Adobe designed its Firefly generative AI model with transparency and disclosure |
| BBC R&D | BBC used face swapping to anonymize interviewees |
| OpenAI | OpenAI is building disclosure into every DALL-E image |
| TikTok | TikTok launched new AI labeling policies to prevent misleading content and empower responsible creation |
| WITNESS | Even the best-intentioned uses of generative AI still need transparency |

**Analysis Methods**

Preliminary, descriptive codes emerged when listening to interview transcripts, followed by more formal inductive coding specifically focused on transparency and disclosure themes (Bingham, 2023). Codes emerged related to the goals of disclosure, qualities of effective and ineffective disclosures, and key tensions and constraints faced by stakeholders. Codes were iteratively organized into higher-level themes for analysis, with a dedicated coding round for the disclosure elicitation and analogical reasoning responses. Codes that emerged from interviews then informed deductive coding of the 13 PAI case studies, which took shape as another rich dataset of expert perspectives to be placed in dialogue with the individual, anonymous expert interviews (Saldaña, 2009).

## Findings and Analysis

Analysis revealed how stakeholders navigate synthetic media disclosure design through three interconnected dimensions. First, they articulate varied goals for effective synthetic media disclosure. Second, they face key tensions when pursuing their disclosure goals under real-world constraints. Third, analogical reasoning supports tension navigation, but not removal. Here, I examine how stakeholders define disclosure goals and effectiveness, the key tensions they confront, and how 36 analogies shed light on transparency decisions. I draw on both interviews and case study analysis for assessing goals and tensions but focus attention on the elicitation task segment of the interview data to analyze analogies.

### What is "Effective" Synthetic Media Disclosure?

Interviews and case studies revealed how those crafting synthetic media understand the goals of AI disclosure. Two key goals emerged that largely align with Wittenberg and colleagues' goals (2024): 1) process transparency—providing factual information about media creation to support independent audience assessment, and 2) harm reduction—mitigating societal and individual risks. However, the harm reduction goal manifested in strikingly varied forms across stakeholder contexts, complicating any simple two-goal model.

#### *The Process Transparency Goal: Disclosure as Information*

Process transparency emphasizes disclosure's purpose as providing factual, process-related details about media *without* judging credibility. It was dominant amongst several policymakers and technology stakeholders, like PAI3 who emphasized their support for helping audiences "understand what [they] are looking at." In their framing, process transparency fulfills that right by clarifying how content was made, *not* by judging credibility.

Microsoft and Adobe both illustrated process transparency goals in their PAI case studies. Microsoft sought to display "the origin of the content, such as the camera model or AI tool used," while Adobe aimed to provide signals about the "origins and changes made to digital files," thereby "giving viewers of digital content important context to help them understand what they are seeing and allowing them to make their own decisions about whether to trust the content." This framework positions disclosure as empowering user autonomy: platforms provide neutral, factual information and users determine significance.

#### *The Harm Reduction Goal: Disclosure as Protection*

A second goal treats disclosure as more than merely neutral information, but rather as an intervention explicitly aimed at preventing harm. Several stakeholders suggested that the process transparency goal can lead to harm reduction. TikTok emphasized how process transparency can "provide context for viewers that reduces the risks of being harmfully misled by [synthetic media] and improves trust and transparency about how content is being made." However, several stakeholders differentiated between the process and harm reduction goals. And harm reduction surfaced in different ways, serving as an umbrella framework encompassing distinct concerns affecting varied disclosure goals and audiences.

**Epistemic Integrity and Media Institution Protection.** News organizations and several civil society groups emphasized disclosure's role in society's capacity to distinguish fact from fiction. Code for Africa, a nonprofit working with African journalists, suggested disclosures can "preempt attacks on content credibility" and "allay mistrust in media," a much firmer, more defensive posture than those merely describing disclosures as "providing transparency around content that was created using AI" (Meta). Here, disclosure is framed as an inoculation against accusations of manipulation by media organizations, preserving institutional credibility and, by extension, the news ecosystem's capacity to anchor shared reality and epistemic integrity.

**Individual and Creator Protection.** Adobe, TikTok, and human rights organization WITNESS emphasized disclosure's role protecting individual creators. Disclosure enables attribution, creative uses of synthetic media, and prevents impersonation or unauthorized uses of one's work and likeness. Here, harm reduction focuses on individual rights rather than overall epistemic deception per se, framing disclosure's harm reduction goals around credit, consent, and control. Importantly, organizations like WITNESS also have a mission focused on other challenges, like broader epistemic integrity.

**Child Safety.** Child safety stakeholders offered another lens for disclosure's harm reduction goals. Rather than informing public audiences or AI-generated child sexual abuse material (AIG-CSAM) consumers, Stanford HAI emphasized disclosure's limited role in harm reduction for child safety overall, and the need to reach "trust and safety teams at online platforms" or "the justice system." Moreover, child safety stakeholders emphasized that AIG-CSAM creators actively avoid disclosure to maximize perceived authenticity. This dynamic then inverts typical disclosure logic, as an effective system requires that process transparency functions despite adversarial resistance. Here, disclosure serves professional investigators in an adversarial context, thereby affecting disclosure design considerations.

### A Fundamental Challenge: Process Transparency is not Enough to Support Media Credibility Judgements

Across both frameworks, stakeholders highlighted a logical challenge that complicates disclosure design: knowing media is AI-generated via process transparency does not indicate whether it is credible or harmful. Conversely, media untouched by AI is not inherently credible and harmless. As POL10 stated, "there's a difference between knowing whether something is synthetic or real and knowing whether it's harmful or not." Yet, while stakeholders recognized that process transparency is an imperfect proxy for credibility and harm, many also emphasized disclosure goals focused primarily on communicating AI involvement. This paradox reflects competing tensions in disclosure design: whether to be normative or neutral, and whether to prioritize proactive warnings or precise information.

### Key Tensions Affecting Disclosure Design and Implementation

Across technology, media, policy, and civil society interviews and case studies, stakeholders described how these competing demands shape disclosure decisions. Understanding how stakeholders navigate them reveals the practical limits and possibilities of disclosure for supporting media credibility assessments.

#### *Tension Between Normativity and Neutrality*

Nearly all stakeholders emphasized their commitment to "neutral" disclosure—providing objective signals about AI's presence without judging media credibility. POL2 described their legislative approach as creating "a bill that helps people identify what's real and what's fake online, not taking a position on the validity or the right or wrong of that content."

However, the aspiration towards neutrality conflicts with articulated harm reduction goals. Stakeholders expressed deep skepticism about whether neutral labels could meaningfully support user understanding. POL10 captured this uncertainty: "So you've applied a label. How does a user make sense of that?" PAI8 continued, "[Labels] might be great for people who are curious, but not for people who probably would benefit the most from knowing the fact that they're looking at AI shit." Others challenged this user-centric approach directly, stating "Why the heck are you pushing the effort down to the users?" (POL9). Code for Africa made the critique explicit: "We are not sanguine about the effectiveness of media literacy education, synthetic or otherwise. In our view, expecting citizens to shoulder the burden of identifying misinformation or synthetic media is a form of victim blaming."

While stakeholders gravitate toward neutral signals, they express how these signals may either not be enough to support harm reduction and may even contribute to greater harm and via faulty media credibility judgments. Google suggests that disclosures may cause harm, stating "we are concerned about unintended implications on user perceptions of content accuracy and this is why we recommend exercising caution when prominently displaying…disclosures." In essence, by placing the evaluative burden on stakeholders through neutrality, stakeholders may be sacrificing their ultimate harm reduction goals. And yet, stakeholders claim success. Companies tout "you know, we're going to fix the trust problem and all that. We'll show you the truth. Even though everyone in their teams are saying don't talk about the truth. You know they can't resist" (PAI1).

Recognizing the inadequacy of purely neutral, process transparency disclosure, stakeholders sometimes seek compromises to provide richer context without blatant normative claims. They do this in two main ways: 1) by making domain-specific compromises and 2) through normative embedding within design choices. Multiple stakeholders highlighted domains where AI use might be more consequential, particularly "broader civics and elections" contexts. Choosing these contexts serves, in essence, as a normative decision about which categories are most consequential for mitigating harm. PAI9 proposed another context-related proxy for normativity suggesting "the difference between a mechanical change and a human change is where the difference of materialities could start to come up."

Others infused their "neutral" transparency mechanisms with normativity through interface design. Google noted, "sometimes, prominent disclosures about the method of content generation may be particularly helpful to users; other times, they may be unhelpful or confusing" while Microsoft described how user research on process transparency UX informed normative decisions about how to present such information. Indeed, even ostensibly neutral signals are laden with normative elements from those implementing them. Whether such compromises reduce the interpretive burden for users remains unclear. Nonetheless, the neutrality-normativity tension fundamentally shapes disclosure design, revealing how stakeholders must balance their commitment to user autonomy with their recognition that neutral signals alone cannot achieve harm reduction goals.

### ***Tension Between <u>Proactivity and Precision</u>***

Stakeholders consistently described another tradeoff: between choosing to proactively and comprehensively label media (thereby accepting false positives) and precisely and accurately applying labels (thereby accepting false negatives). Several platforms and policymakers, responding in a "super election year," justified a proactive disclosure stance due to pressures from imminent elections and AI's novelty. As Meta wrote, "2024 saw a historic number of elections around the world, and we deemed it important to prioritize greater transparency, which made us more comfortable with potentially overlabeling." In the media context, Code for Africa emphasized a commitment to proactively labeling all content, to "inoculate news organizations against accusations of manipulating information, and to give their audiences the comfort of knowing that there are fixed, accessible editorial standards and policies governing the news organizations' content production."

Yet, in practice, this proactive approach may not serve disclosure's harm reduction goals. When Meta relied on technical signals to proactively label media as "Made with AI," this resulted in media with minor tweaks being labeled, providing transparency that did not empower audiences to make meaningful judgments. Complicating their support for proactivity and illuminating the tradeoff, Code for Africa explained "excessive or vague labeling may scare off individuals with already low trust...But failing to provide audiences with information they may want to decide what news to use and trust could equally prove damaging." These instances illustrate how overlabeling may erode, rather than enhance, users' ability to make meaningful credibility judgments that mitigate harm, but so too might not proactively labeling.

Other user testing highlights the benefits of the precision imperative, showing how overlabeling may not simply be helpful or benign, but could be harmful to media credibility goals. Google reported that overlabeling could trigger implied truth and authenticity effects. PAI10 confirmed this perspective, saying "we haven't had a realistic conversation [about what needs to be labeled] which now means that we've kind of lost people because now they're like oh, everything gets labeled and it shouldn't, and it's annoying and contributes to the implied truth effect." Indeed, the precision vs. proactivity debate is not simply about technical accuracy of the labels, but also about supporting user understanding, perceptions, and the impact of the disclosures.

Several stakeholders articulated attempted compromises between proactivity and precision. Some accepted proactivity for now to move the field towards eventual precision, stating that disclosure regimes could change over time to accommodate shifting user needs. PAI12 advocated for starting with proactivity to build user expectations over time, stating "it's not about making sure that everything's completely robust…it's about making it common enough for people to see over time that they have their own internal red flags that something feels off." Another compromise involved acceptance and accountability, emphasizing that both strategic poles—whether full proactivity that overlabels or precision that leads to underlabeling—can cause challenges. Accepting this requires stakeholders to focus on how these risks may "disproportionately affect communities that are already underserved or under-represented in technology development and infrastructure" (Meedan). The focus then, for designers, may not just be dwelling on the proactivity and precision trade-off, but designing complementary systems that "include opportunities for appeal" for labels, especially as "existing appeals processes for reporting [are] limited."

The proactivity-precision tension thus shapes what gets labeled and how, but it also mediates how stakeholders conceptualize the actual purpose of disclosure. Is disclosure ultimately a protective system that accepts imprecision for emergent AI literacy and user safety, or is it an intervention that supports accuracy to best enable audience agency? Interestingly, both solutions can place the interpretive burden on audiences to develop their own heuristics for when disclosures ultimately matter, potentially disadvantaging those without existing media literacy, technical knowledge, or more simply, time.

### ***Connections Between the Two Tensions***

While analytically distinct, these two tensions—between normativity and neutrality and proactivity and precision—interact in practice. Stakeholders' embrace of proactive labeling can serve as a strategy to avoid making explicit normative judgments; if everything is labeled, no determination of significance or risk assessment is required. Conversely, achieving precision necessarily requires some form of normative judgment about which AI uses warrant disclosure, as Google exemplifies through its decision to inform disclosures based on media "sensitivity." This interaction reveals that stated preferences for "neutral" disclosure may be achievable only through comprehensiveness that renders labels less informative, while informative labels that may place less of the interpretive burden on audiences inevitably encode normative assumptions about what matters to them. These interacting trade-offs reveal that effectiveness in disclosure design is not a purely technical matter, but a sociopolitical balancing act between interpretive responsibility and institutional accountability.

### Disclosure Analogies Inform Disclosure Design

This balancing act, along with the lack of empirical consensus from user studies, requires stakeholders turn to additional conceptual tools to make disclosure decisions. Analogies serve as one such tool, helping them reason about what disclosure should resemble, drawing on other governance or communication frameworks that have harm reduction goals to address synthetic media's challenges.

23 interviewees offered a total of 36 analogies for disclosures from other domains (27 of which were distinct). Unsurprisingly, the **nutrition label** was mentioned the most (by seven participants). Only two other analogies repeated across stakeholders: **California's Proposition 65** (**Prop 65**) which requires warnings for products and places containing chemicals known to cause cancer, birth defects, or reproductive harms, mentioned by three participants, and **passports**, described by two participants. Of the total list, most analogies were from consumer protection; ten analogies focused on food-related information (including nutrition label references), seven focused on toxicity and public health (including those on Prop 65 but also others like **European cigarette labels** and **makeup toxicity scores**), and six invoked cybersecurity and privacy disclosures (such as **General Data Protection Regulation or GDPR** from Europe or other more visual disclosures like the **HTTPS padlock** or **SSL security signals**). Less pronounced themes included animal welfare (**egg quality ratings** and **RSPCA**), **energy efficiency**, entertainment, identity verification, and even household instructions. Taken together, these analogies help probe how decision makers navigate the complex goals and tensions surrounding synthetic media disclosure.

#### ***Positive and Negative Characteristics of Disclosure Analogies***

Analogies revealed positive and negative features to apply to synthetic media disclosures. Of the 36 total analogies referenced, 21 were described as effective, 10 as ineffective, and five as both **(Table 3).** Notably, stakeholders often evaluated the same analogies differently. Of the seven references to nutrition labels, two described them as ineffective and five as effective. Prop 65 was described as effective by one stakeholder, ineffective by another, and as both by a third. The two interviewees who referenced passports both implied they effectively communicate identity-related policy goals.

Of course, how stakeholders assess an analogy can depend on what problem they believe disclosure should solve. However, for synthetic media disclosure design, what matters most is not whether these analogies are objectively effective in their original domains, but rather how stakeholders transpose their understanding of these mechanisms—and their perceived successes or failures—onto the AI context.

Table 3: Disclosure Analogies Described as Effective, Ineffective, and Both ($N$ = 36)

| Effective Disclosures | Ineffective Disclosures | Both Ineffective + Effective |
|---|---|---|
| Recipe | European Cigarette Warnings | California's Prop 65 |
| Nutrition Label ($n$ = 5) * | Sugar Labels | Blue Tick (like Twitter) |
| UL Safety Marks | GDPR | SSL + Lock Icons |
| Historical Movie Labels | U.S. Surgeon General Warnings | HTTPS Padlock |
| Passports ($n$ = 2) * | Redline Overlay from the New York Times | German Egg Quality of Life Rating |
| Laundry Washing Labels | Nutrition Label ($n$ = 2) * | |
| California's Prop 65 | RSPCA (for animals in the UK) | |
| Apple's Privacy Scores | Medication Side Effects | |
| Makeup Toxicity Scores | California's Prop 65 | |
| Explicit Movie and Music Ratings | | |
| Energy Star Label (specific) | | |
| Energy Efficiency Label (general) | | |
| U.S. Cyber Trust Mark | | |
| Bike Serial Number | | |
| Political Disclosures | | |
| Organic Label | | |

**Note:* $N$ denotes the frequency of occurrence within the sample. Acronyms: GDPR (General Data Protection Regulation), RSPCA (Royal Society for the Prevention of Cruelty to Animals), UL (Underwriters Laboratories)

### *What Not to Do: Lessons from Nutrition Labels*

The nutrition label was the most frequently referenced analogy, appearing seven times across interviews. Interestingly, however, stakeholders not only focused on what makes nutrition labels effective—as public discourse about AI disclosure often does—but also on their limitations. PAI1 explained they are "very opposed to the nutrition label metaphor," going on to say "it's really not about nutrition, whether [media] is good or bad for you and it's also not about like, a fixed set of ingredients, right?" Through such critique, PAI1 helps navigate the normative vs. neutral tension, emphasizing that "good or bad," "healthy or not" is not the right frame for synthetic media disclosure. This stakeholder is more supportive of an alternative food analogy, the **recipe**, because with media, "you sprinkle something else in, you cook it, then you put it in your freezer, and then you break down and reheat it." Synthetic media disclosures, unlike nutrition labels, should describe a process rather than static characteristics.

Another nutrition label critique focused on how they can be manipulated to obscure harms. As PAI5 highlighted, "we get told [they] have reduced sugar and you go oh great and actually it's got loads more fat and it's just, like, full of aspartame, or whatever. So basically companies will use these labels in a deceptive way." The nutrition label becomes a constraining frame for synthetic media disclosure because it may not ultimately shift corporate behavior to mitigate harms, whether epistemic or health related.

A third critique focused on user interpretability. POL3 questioned whether nutrition labels actually support user understanding: "how much do people really understand about nutrition labels, like even as a highly educated person? There's pieces of nutrition labels where I'm like, percentage zinc, yep, okay, what am I supposed to do with that?" The nutrition label analogy thus reveals that effective disclosures must do more than merely provide information to promote physical or epistemic health. They must identify processes, mitigate unintended harms, and ensure users can actually interpret them.

*Disclosures Can Simultaneously be Effective and Ineffective via Different Pathways*

Most stakeholders highlighted their analogies as either explicitly negative examples or positive examples for synthetic media disclosure. However, some disclosures were perceived as both effective and ineffective by the same stakeholder, illustrating how a single mechanism can operate through multiple pathways. PAI3's Prop 65 commentary is particularly illustrative, as they describe how initially it was effective by changing corporate behavior:

> *"I don't think shame is really, like, a useful industrial concept, but what may surprise you is that if you're going to have fewer customers [after Prop 65 transparency], and a consumer has the option of one that has, you know, lead in it, and you need to tell them and one that does not have that, people are going to go to your competitor, so why don't you just reformulate?"* (PAI3)

In this view, transparency functioned as a competitive pressure, prompting companies to remove toxins rather than risk losing customers. However, PAI3 then noted that the transparency mandate has become too broadly applied, diluting its effectiveness

> *"It did change a number of products for the safer, like probably a non-negligible number of people don't have cancer because of it, but now we have signs in front of Disneyland parking lots that give Prop 65 warnings upon driving in…"* (PAI3)

Initially, disclosure prompted meaningful corporate behavior changes that reduced harm. Over time, however, overuse and oversaturation diluted its impact, rendering warnings less meaningful. For PAI3, this trajectory is "particularly instructive in the synthetic media side of things"—suggesting AI disclosure requires not just implementation, but also careful scoping and monitoring over time to maintain relevance and impact.

*Navigating Corporate Incentives to Support Standardization and Cooperation*

As stakeholders further reflected on their analogies, many emphasized the need to navigate corporate incentives and push towards standardization for synthetic media disclosures to be effective.

**Navigating Corporate Incentives.** Several stakeholders warned that without careful design and enforcement, disclosures risk serving corporate rather than consumer interests. As PAI5 cautioned, "companies will learn to use these labels in a deceptive way," leading to "AI disclosure washing." Similarly, POL5 compared the dynamic to **medical disclaimers,** where pharmaceutical companies "check the box" on side-effect warnings rather than offering meaningful information. PAI3's discussion of Prop 65 further illustrated how competitive pressures can initially drive effective disclosure but ultimately incentivize defensive overlabeling that renders warnings meaningless to consumers.

**Standardization.** Several stakeholders identified standardization as essential to counteracting these corporate pressures. PAI10's analogy focused on social media's "**blue tick**" verification system on social media, viewing it as successful precisely because "it emerged very quickly and normally competitive platforms all took it up in the same way." **Passports** provide another useful standardization example, "All passports have common functionality, common fields, and then common ways to be read as you cross borders, in different countries" (PAI9). The analogies highlight how standardization can support disclosure, complementing responses in case studies; organizations emphasized how standardization could help both those disclosing, and those consuming, disclosures interpret them effectively. Standardization creates a level playing field that reduces incentives for disclosure washing while increasing user comprehension through consistent signals.

*The Epistemic Dimension: What Synthetic Media Adds to Traditional Disclosure Frameworks*

Beyond lessons about standardization and corporate incentives that help manage tensions and promote design directions, the disclosure elicitation task also revealed additional conflicts. Stakeholders identified characteristics of effective disclosure that are difficult, and potentially impossible, to achieve simultaneously. Some emphasized that disclosures should be authoritative—like **energy ratings** or **passport** iconography backed by government agencies—yet also allow people to make their own decisions based on neutral information. Others argued disclosures should make harms extremely salient, like **European cigarette warnings** featuring photographs of diseased lungs, while others praised disclosures as simply effective through their absence, like **SSL locks** in cybersecurity where only the missing signal alerts users to risk. These contradictions may stem from the fundamental limitations of an analogy that is not exact, and the epistemic complexities of synthetic media risks. As POL3 explained:

> "*I am generally very wary about any of the analogies like the whole, I don't know, misinformation was the new smoking, is so weird. I think none of the analogies really capture the state of the problem. And that also reveals the deeper problem of, you can't really fit the issue into a platitude, and that also makes it much more difficult to understand. And then if it's much more difficult to understand then it's much more difficult for people to, like a layman, to understand it and interpret it and to understand how it affects their lives"* (POL3).

Unlike nutrition labels or cigarette warnings, synthetic media disclosures cannot escape questions of truth and authenticity that are inherently value-laden and context-dependent (Leibowicz, 2021). While analogies like nutrition labels are also normatively informed, the values they convey are relatively standardized—most societies agree that high sugar is bad or smoking causes cancer. By contrast, what counts as credible or harmful synthetic media is often subjective, socially negotiated, and tied to evolving norms around truth, trust, and credibility.

Nonetheless, these analogies remain instructive because they help stakeholders identify and articulate additional dimensions of disclosure and work through inherent tensions—negotiating what they are comfortable with and making explicit the trade-offs involved. Recognizing these limits helps explain why tensions between neutrality and normativity, and between comprehensiveness and precision, are so difficult to resolve in practice.

## Discussion

As synthetic media evolves—enabling anyone to generate photorealistic videos of political candidates or resurrect historical figures—the effectiveness of disclosure-based governance remains uncertain. What is certain, however, is that policymakers and practitioners designing disclosures operate under sociotechnical constraints, and how they navigate these tensions ultimately impacts billions of media consumers. This research therefore shifted the unit of analysis to upstream decision makers, illuminating several key themes in how they translate transparency goals into policy mandates: 1) tension management rather than tension resolution, 2) a mismatch between central analogies for epistemic challenges, and 3) convergent disclosure design practices.

### Analogical Reasoning as Tension Management, not Resolution

The use of analogies as frames to manage, though not fully resolve, tensions, emerged throughout almost all interviews, and tension alleviation was emphasized in case studies, too. Stakeholders frequently turned to consumer protection analogies like nutrition labels and energy efficiency warnings not to completely alleviate trade-offs, but as frames and scaffolds to approach and communicate design conflicts. Yet, despite recognizing the limitations of existing disclosure mechanisms from analogous fields, stakeholders often implemented synthetic media disclosures that replicated many core features of those analogies.

Disclosure implementation may therefore seem paradoxical. Stakeholders invoke analogies of neutral labels that place unfair interpretive burdens on users, yet build neutral, process transparency-oriented labeling systems at the same time. They recognize proactive labeling can risk desensitizing audiences yet choose to prioritize greater transparency in moments of political intensity and technological integration. They understand that the presence of AI may not indicate harm, just as a Prop 65 label on a couch may not accurately indicate harm, yet they build disclosure systems centered on signaling AI involvement. These paradoxes likely arise because tension management through analogical reasoning does not fully resolve underlying conflicts, particularly those shaped by complex sociopolitical forces such as free expression, censorship, and market dynamics (DiResta, 2025). Instead, analogies make tensions navigable while simultaneously encoding solutions—neutral signals, user interpretation, comprehensive labeling—that stakeholders themselves may question. Beyond alleviating tensions, institutions across media, industry, and other sectors may require evocative analogies to transcend competitive dynamics that could undermine disclosure effectiveness and standardization. Analogies like the nutrition label, while both constraining and supportive, provide a shared, concrete language that stakeholders can use collaboratively to move toward more standardized disclosure systems.

### Moving Beyond Consumer Protection Analogies for Epistemic Problems

Several stakeholders, notably POL3, acknowledged that analogies may have limited capacity to address key tensions because of the epistemic challenges posed by synthetic media. When dealing with disclosures about media authenticity, which hinge on both content and context, moving beyond consumer protection frames could better support tension alleviation for disclosure stakeholders. Analogies drawn from visually or epistemically oriented contexts may offer more effective frameworks for resolving tensions in synthetic media disclosure design.

For example, disclosure and transparency schemas used by journalists, visual archivists, and museum curators may better align with the epistemic realities of synthetic media disclosure design. Photo editors in newsrooms, for instance, have decades of established norms for determining what counts as a material photo edit. Museums often go beyond labeling artifact provenance, providing interpretations that acknowledge contested meanings. Importantly, these professions treat credibility assessment as a process requiring institutional intermediaries and professional judgment—not solely audience interpretation—while also navigating questions of who counts as a qualified professional. For instance, in 2021, the New-York Historical Society allowed labels to be written by public intermediaries, not just curators; a painting of a horse-drawn carriage, for example, was contextualized by a real-world carriage driver (Jacobs, 2021). Drawing analogies from fields that actively manage interpretive burdens may empower stakeholders to adopt more of this curatorial role in synthetic media disclosures, while avoiding censorship. Stakeholders already confront aspects of this role, making domain-specific compromises—for example, around elections—and designing user interfaces that better enable audience credibility assessments. Scaling the protocols of a museum or newsroom to the speed and volume of online media is undoubtedly challenging. Nevertheless, analogies that emphasize visual material and strategies for meaningfully stewarding trust and truth may more effectively guide disclosure design decisions.

### Emergent Disclosure Best Practices

Despite the imperfection of the analogies and complexity of disclosure tensions, stakeholders did offer an emerging set of best practices that inform their disclosure design and decision making. Effective disclosures should: be trusted, visually standardized, have enforcement mechanisms, be aesthetically pleasing, be straightforward, difficult to remove, and intuitive, telegraph only what is necessary information, be unobtrusive, and solely describe material changes. While many of these prove difficult to pursue in practice, a challenge which has been the focus of this paper, stakeholders do have an aspirational sense of how disclosures may look.

## Limitations and Future Work

As noted in the literature review, this study focuses on decision-makers rather than end-users to understand the upstream dynamics shaping AI disclosures. However, this approach cannot replace future research directly testing and centering user interpretation in media credibility research. Any user understanding presented here is indirect—filtered through stakeholders' interpretations—and helps scholars form inferences rather than empirically verify user comprehension. Future studies must examine user needs directly, including the gap between stated preferences for labels and actual attitudes when encountering them, the realities of implied truth and authenticity effects, and how user understanding evolves through longitudinal, real-world data rather than laboratory settings.

In addition, the North American and European focus of this work limits broader claims about disclosure decision-making. While North America and Europe are leaders in the disclosure implementation and rulemaking, they present a specifically Western viewpoint on how AI and content is governed. China, who has also been spearheading direct disclosure legislation, may likely have very different concerns when thinking about the neutrality and normativity tradeoff, for instance, since they are operating in an information environment accustomed to speech restrictions and censorship. Future research should include policymakers and media, industry, and civil society stakeholders from beyond these two continents. Additionally, expanding beyond qualitative methods to experiments or surveys might not only capture broader, more global samples, but also address the generalizability limitations affecting qualitative studies like this one.

## Conclusion

Ultimately, this paper untangles how the AI policymakers and practitioners designing synthetic media disclosures navigate complex goals and transparency implementation realities. Through interviews and case reporting across civil society, policy, industry, and media sectors, this study shows how real-world disclosure stakeholders navigate tensions between normativity and neutrality as well as proactivity and precision. Analogical reasoning (from nutrition labels to toxicity warnings) both enables and constrains disclosure design, serving as a tool but not a solution for navigating key tensions. By studying the stakeholders designing disclosures, this paper can inform scholarship and real-world decision making in service of audience empowerment in the increasingly AI-infused media environment.